\documentstyle[psfig,11pt,fullpage]{article}
\begin{document}

\begin{center}
\thispagestyle{empty}
{\Large{\bf Modifications of  The $BTZ$ Black Hole by  
 a Dilaton/Scalar}}\\ 
\vspace{1.5cm}
Kevin C.K. Chan\footnote
{\normalsize Email address: kckchan@avatar.uwaterloo.ca}\\
Department of Physics, University of Waterloo,\\
 Waterloo, Ontario, Canada N2L 3G1.\\
\end{center}
\vspace{0.5cm}
\centerline{ABSTRACT}
\bigskip

We investigate some modifications of the static $BTZ$ black hole solution due to a chosen asymptotically constant dilaton/scalar. New classes of static black hole solutions are obtained.
One of the solutions contains the Martinez-Zanelli conformal black hole solution as a special case. Using quasilocal formalism, we calculate their mass for a finite spatial region that contains the black hole. Their temperatures are also computed. 
Finally, using some of the curvature singularities
as examples, we investigate whether a quantum particle behaves singularly or not.

\bigskip

\noindent PACS number(s): 04.70.Bw, 04.50.th, 64.60.Kz


\section{Introduction}

Studies of lower dimensional gravitational theories continue to attract the
attention of theorists for a number of reasons.
It is always of interest to see how the dimensionality
of spacetime affects the physical consequences of a given theory. 
For example,
static spherically symmetric ($SSS$) vacuum solutions of general relativity
($GR$) in four or higher dimensions are Schwarzschild black holes. However,
the same set of vacuum field equations yields a locally flat conical spacetime
in three dimensions \cite{dhj}.
In other contexts, lower dimensional theories of gravity are widely used as
``toy models'' for studying varies classical and quantum aspects of gravitation
which are conceptually and mathematically difficult to handle in 
four or higher dimensions. 

Many lower dimensional gravitational theories are not without consequence:
For example, they admit non-trivial black hole or cosmological solutions
(see, {\sl e.g.} \cite{mann} for a review). The black hole
conundrum has long been one of the
most outstanding problems of modern physics. It has remained in focus
as one of the potential testing grounds for quantum-gravitational
phenomena for a long time.  It is desirable
to search for new black hole solutions in any spacetime dimensions.

We consider here dilaton and scalar-tensor ($ST$) theories as alternative theories of gravity in three spacetime dimensions. As previously mentioned
three dimensional vacuum $GR$ admits no black hole
but rather a trivial locally flat (globally conical) solution.
One has to either couple matter to $GR$ ({\sl e.g.} a cosmological
constant \cite{btz} or dilaton as matter \cite{chaman}),
or consider alternative vacuum (or non-vacuum)
gravitational theories in order to get black hole solutions.
Motivated by this, we  search for new black hole solutions
in $GR$ coupled to a self-interacting dilaton and vacuum $ST$ theories.
In particular, we will see that these black holes are modifications of the  $BTZ$ black hole by an asymptotically constant dilaton/scalar.

In three dimensions, dilaton and $ST$ black holes
were  obtained in \cite{chaman} and in \cite{lemos, sa} respectively.
The most general action
coupled to a dilaton/scalar can be written as~\cite{luis}
\begin{equation}
S=\int  d^3x {\sqrt{-g}} [C(\phi){\cal R}-{\omega(\phi)}(\nabla\phi)^2
+V(\phi)], \label{act}
\end{equation}
where ${\cal R}$ is the scalar curvature,  and $V(\phi)$ is a potential function for
$\phi$. $C(\phi)$ and $\omega(\phi)$ are collectively
known as the coupling functions.
For simplicity, we have restricted ourselves to one dilaton/scalar
$\phi$ only. It is worthwhile noting  that
the two dimensional version of (\ref{act}) and
generic black hole solutions of its field equations
have attracted interest for a period of time (see,
{\sl e.g.} \cite{kunstatter}).

Special cases to (\ref{act}) in three dimensions were previously considered by a number of authors.
The first example is the static  $BTZ$ black hole solution to $C(\phi)=1$,
$\omega(\phi)=0$ and $V(\phi)=2{\Lambda}$ \cite{btz}. The second
example corresponds to the same
$C(\phi)$ as above, but with a non-trivial $\phi$, $\omega(\phi)=4$
and $V(\phi)=2{\Lambda}e^{b\phi}$, for which
static  black hole solutions have been previously derived  in \cite{chaman}.
These examples have the condition $C(\phi)=1$, for which the metric coupling to matter
is the Einstein metric.  This is no longer true when $C(\phi)\neq 1$, and the gravitational
force is governed by a mixture of the metric and scalar fields;
the class of actions (\ref{act}) with a nontrivial $C(\phi)$ are called $ST$ theories.

Without loss of generality, $C(\phi)$ can be normalized to $\phi$ by a scalar field redefinition.
Obviously, one can carry out  a conformal transformation on the metric to (\ref{act}) such that $C(\phi)$ is ``rescaled'' to $1$, and the resulting action is
three dimensional $GR$ coupled to dilaton matter \cite{wand}. Adopting the viewpoint that
both scalar and metric fields govern the behaviour of gravity,
(\ref{act}) with $C(\phi)=\phi$ is the natural choice.
Our terminology is as follows: for $C(\phi)=1$, we call (\ref{act}) 
dilaton gravity ($GR$ coupled to a dilaton); otherwise it is referred to as the
$ST$ gravity if $C(\phi)=\phi$.
The $ST$ gravity  can be interpreted as
vacuum gravity \cite{luis}. Thus any solution to it may be called
a vacuum solution. In general, one may add a matter action to (\ref{act}).
For example, Verbin constructed a three-dimensional scalar-tensor gravity in \cite{verbin}.
It is dimensionally reduced from four dimensional $GR$ coupled
to matter. The resultant action becomes (\ref{act}) with $\omega(\phi)=V(\phi)=0$
but with an additional matter Lagrangian coupled to the scalar \cite{verbin}.

The third example, corresponding to a static, horizon-free solution to the choice $\omega(\phi)={{\omega}_o\over \phi}$ and $V(\phi)=0$, was obtained in \cite{romero}.
This is a solution to the three-dimensional 
Brans-Dicke theory without a potential term. The fourth example
is a three-dimensional $ST$ black hole with the choice
$\omega(\phi)=0$ and $V(\phi)=2{\Lambda}$ \cite{lemos}  .
Recently, it was found that static black holes exist in three dimensional
Brans-Dicke theories with a variety of Brans-Dicke parameters
and a potential term $V(\phi)=4{\lambda}^2\phi$ \cite{sa}; in particular,
when $\omega(\phi)=-{1\over \phi}$, the action becomes the simplest low energy string action.
The only black hole solution to this string action is trivially obtained through
the product of the two dimensional black hole with ${\bf{S}}^1$
\cite{chaman, sa}.  A review of some of the above-mentioned black hole solutions may be found in \cite{carlip}.  More recently, a  three-dimensional black hole solution in the conformal coupling   case ($C(\phi)=1-{1\over 8}\phi^2$, $\omega=1$ and $V=2\Lambda$) was reported in \cite{ZAN}. In section 5, we will show that it is a special case of one of the solutions obtained in this paper. In  other contexts, investigations of semi-classical properties of a massless  conformally coupled scalar field  in the $BTZ$ background were carried out  in \cite{semi}.  

We consider in this paper the extraction of new static circularly symmetric ($SCS$)
solutions for the field equations derived from (\ref{act}).  We will see that these solutions are modifications of the $BTZ$ spacetime in the sense that when the asymptotically constant dilaton/scalar  is switched off, they all reduce to the $BTZ$ case. It is impossible to get the most general exact solutions
in terms of arbitrary $\omega(\phi)$ and $V(\phi)$, unlike the
two-dimensional cases studied in  \cite{kunstatter}.
One needs, for whatever motivation, to specify the choices for
the coupling and potential functions, and then solve for exact
solutions. Each possible choice of those coupling and potential functions
may have interesting solutions. Here we do not attempt to study the whole
set of solutions of different possible choices. Rather,
our potential functions are motivated by the forms previously considered in
two and four dimensions.

The field equations for $C(\phi)=1$ and $C(\phi)=\phi$
will be derived in section 2.  In section 3, we present the equations of 
quasilocal formalism developed in \cite{broyor, jolien} for  calculating mass for a finite spatial
region that contains the black hole. 
In section 4, we consider the dilaton gravity case
and get three exact solutions for three different
choices of the potential function $V(\phi)$.
One solution has a naked timelike
singularity, and the other two  are black hole solutions
with a ring curvature singularity
({\sl i.e.}, the Ricci scalar diverges at a finite circumference).
We then investigate $ST$ gravity in section 5.
We get three exact black hole solutions for three sets
of choices of coupling and potential functions. 
One of the solutions contains the Martinez-Zanelli conformal black hole as a special case. In section 6, we compute the entropy
and the temperature of these black holes.  Finally, using the
naked and ring singularities in the dilaton gravity as examples, we investigate
whether or not a quantum particle behaves singularly in such spacetimes,
following the approach in \cite{hm}. We summarize our results in the concluding section.
Our conventions are as in Wald \cite{wald}.

\section{Field Equations}

We first derive the field equations for action (\ref{act}) with the choice
$C(\phi)=1$ and $\omega(\phi)=4$. Varying (\ref{act}) with respect to
the metric and dilaton fields, respectively,
yields (after some manipulation)
\begin{equation}
{\cal R}_{\mu\nu} =  4\nabla_{\mu}\phi\nabla_{\nu}\phi -
g_{\mu\nu} V,  \label{eoma} 
\end{equation}
\begin{equation}
8\nabla^2\phi + {dV \over d\phi}  =  0. \label{eomb} 
\end{equation}
This set of field equations was considered in \cite{chaman} for an exponential
 potential function. We wish to find exact $SCS$ solutions to 
with a different choice of  $V(\phi)$.
In $2+1$ dimensions, the most general such metric has two
degrees of freedom \cite{melvin}, and can be written in the form
\begin{equation}
ds^{2}=-f(r)dt^2+{h(r)\over f(r)}dr^2+r^2d\theta^2. \label{escs}
\end{equation}
With the
metric (\ref{escs}), the equations of
motion (\ref{eoma}) and (\ref{eomb}) are reduced to the following equations: 
\begin{equation}
{f''\over h}-{1\over 2}{f'h'\over h^2}+{f'\over rh} =  2V, \label{eeoma}
\end{equation}
\begin{equation}
h =  exp\left(\int 8 r(\phi')^2dr\right), \label{eeomb}
\end{equation}
\begin{equation}
{1\over h}\left(f' - {h'f\over 2h}\right) =  rV,  \label{eeomc}
\end{equation}
\begin{equation}
{8\over {\sqrt{h}} r}\left({fr\over\sqrt{h}}\phi'\right)'
 + {dV\over d\phi} =  0. \label{eeomd}
\end{equation}
The prime denotes the ordinary derivative with respect to $r$.
${\cal R}_{tt}$ yields (\ref{eeoma}). ${\cal R}_{rr}$ and (\ref{eeoma}) together yield (\ref{eeomb}).
${\cal R}_{\theta\theta}$ gives (\ref{eeomc}).
The last $\phi$ equation (\ref{eeomd}) follows from the local conservation
law $\nabla^{\mu}T_{\mu\nu}=0$. 

We now derive the field equations for $ST$ theories.
Varying (\ref{act}) with respect to the remaining fields yield
\begin{equation}
G_{\mu\nu} =  g_{\mu\nu}{V\over 2\phi} + {\omega\over\phi}\left
(\nabla_{\mu}\phi\nabla_{\nu}\phi - {1\over 2}g_{\mu\nu}(\nabla\phi)^2\right)
+ {1\over\phi}\left(\nabla_{\mu}\nabla_{\nu}\phi - g_{\mu\nu}\nabla^2\phi\right),
\label{eqsta}
\end{equation}
\begin{equation}
2{\omega}\nabla^2\phi + {\omega}(\nabla\phi)^2{d\over d\phi}ln(\omega) + \left({\cal R} + {dV\over d\phi}\right) =  0, \label{eqstb}
\end{equation}
where without loss of generality, we have set $C(\phi)=\phi$ in (\ref{act}).
The local conservation
equation holds due to the Bianchi identity $\nabla_{\mu}G^{\mu\nu}=0$ and it
implies (\ref{eqstb}). We note that one can have
$g_{tt}=-{1\over g_{rr}}$ ({\sl i.e.} $h(r)=1$ in (\ref{escs})) without causing $\phi(r)$ to be trivial.
For simplicity, we set $h(r)=1$. Now using (\ref{escs}) with $h(r)=1$,
(\ref{eqsta}) becomes
\begin{equation}
f'' + {f'\over r} =  2{V\over \phi} - {1\over\phi}\left(f'\phi' + {2\over r}(fr\phi')'\right), \label{eeqsta}
\end{equation}
\begin{equation}
{\phi''} + \omega({\phi'})^2 =  0, \label{eeqstb}
\end{equation}
\begin{equation}
f'r = r^2{V\over \phi} - {1\over\phi}(fr^2\phi')'. \label{eeqstc}
\end{equation}
Similarly, (\ref{eqstb}) can be simplified 
using (\ref{escs}). It is easy to check that the $ST$ solutions in \cite{lemos, sa, romero}
satisfy this system of differential equations.

The $BTZ$  black hole solution to the system of differential equations (\ref{eeoma})-(\ref{eeomd}) and
(\ref{eeqsta})-(\ref{eeqstc}) corresponds to 
$\phi=constant$ and $V=2\Lambda$ which imply $f=-M+\Lambda r^2$ and $h=1$.   If one constructs non-trivial $\phi(r)$ and $V(\phi)$ with the properties such that  $\phi(\infty)\rightarrow constant$ and $V\rightarrow 2\Lambda$, the resultant metrics should asymptotically approach the $BTZ$ one. 

In section 4, we will solve the system of differential equations, (\ref{eeoma})-(\ref{eeomd}). We first assume a special form
for $V(\phi)$ which was previously investigated in other spacetime dimensions.
Then we assume
an asymptotically constant $\phi(r)$ to solve for $h(r)$ from (\ref{eeomb}).
Finally we use $h(r)$, $\phi(r)$ and the chosen $V(\phi)$
in (\ref{eeoma}) and (\ref{eeomc}) to simultaneously solve for $f(r)$.
Note that (\ref{eeomb}) implies that a nontrivial $\phi(r)$ will yield a nontrivial $h(r)$ and vice
versa. One cannot have $g_{tt}=-{1\over g_{rr}}$ when 
$\phi(r)$ is not a constant. Then, in section 5, we will solve (\ref{eeqsta})-(\ref{eeqstc}). Similarly,
we first  assume special forms of $V(\phi)$ and $\phi$. By using (\ref{eeqstb}), 
we obtain $\omega$. Equations (\ref{eeqsta}) and (\ref{eeqstc}) are then solved consistently for $f(r)$. 

\section{Quasilocal Mass}

We will use the quasilocal formalism to compute the mass of
the solutions. If a $SCS$ metric is written in the following form
\begin{equation}
ds^2 =  -f(r)dt^2 + W^{-2}(r)dr^2 + r^2d\theta^2. \label{massa} 
\end{equation}
then the conserved quasilocal mass measured by a static observer at $r$ (the radius of the spacelike hypersurface boundary) for dilaton gravity ($C(\phi)=1$)
is given by \cite{broyor}
\begin{equation} 
{\cal M}(r) =  2\sqrt{f(r)}[W_{0}(r)-W(r)]. \label{massb}
\end{equation}
The terms inside the bracket is identified as the quasilocal energy $E(r)$ which is the thermodynamic internal energy. Thus mass and energy differ by a lapse function. Here $W^2_{0}(r)=g^{rr}_0$ is an background solution which
determines the zero of the energy.
The background solution can be obtained simply by setting constants of
integration of a particular solution to some special value that then
specifies the reference \cite{jolien}. 
When the spacetime is asymptotically flat, the usual $ADM$ mass $M$ is
the mass determined in (\ref{massb}) in the limit $r\rightarrow\infty$.
For solutions with general asymptotic conditions and
without any cosmological horizon or curvature singularity at spatial
infinity, the large $r$ limit of (\ref{massb}) is used to determine the
analogous $ADM$ mass parameter $M={\cal M}(\infty)$.

For $ST$ gravity, the quasilocal mass, according to Hawking and Horowitz prescription, is given by  \cite{jolien, HH, CCM}
\begin{equation} 
{\cal M}(r) =  2r\sqrt{f(r)}\left(\phi'+{\phi\over r}\right)(W_{0}-W). \label{massc}
\end{equation}
We see that if $\phi(r)=1$, then (\ref{massc}) reduces to (\ref{massb}). Both equations
will be used to identify the analogous mass parameter $M$ in our solutions.
Note that the quasilocal mass is not necessarily positive
definite for a spacetime \cite{jolien}. For example,
${\cal M}(r)$ of a given solution may admit negative values for a certain range of
$r$. We will consider that a solution is physical as long as the analogous
$ADM$ mass $M$ is positive.

\section{Exact Solutions in Dilaton Gravity}

We begin by searching for solutions in the dilaton gravity case: $C(\phi)=1$
and $\omega=4$.  We note that potential and/or coupling functions
in terms of a polynomial of a certain function of $\phi$
with unspecified coefficients are not uncommon among field theories;
for example, the four-dimensional bosonic action contains a coupling
function in terms of a power series: $\sum_n c_ne^{2nk\phi}$,
and the coefficients $c_n$ are presently unknown \cite{bento}.
There are a number of studies  in two and four dimensions (see,
{\sl e.g.} \cite{ellis, trodden, chanmann, bech}) in which the authors search for the forms of the potential/coupling functions which yield desired exact solutions.
In particular, the idea of deriving the inflation potential in four dimensions by imposing
particular theoretical or phenomenological requirements has been widely
discussed in recent literature ({\sl e.g.} \cite{adams}).
In other contexts, for example, parameters in certain four-dimensional
theories in early universe are unknown so that they can be adjusted to
avoid certain cosmological  difficulties such as the domain wall problem \cite{dvali}.
All the above-mentioned {\sl ad-hoc} approaches are justified by the fact that
there are no {\sl a-priori} underlying principles that uniquely
specify the potential or coupling functions. Here we use the similar approach to obtain three-dimensional black hole solutions with an asymptotically constant dilaton/scalar.
Motivated by this, we generalize our potential as  a polynomial in certain function
with unspecified coefficients which can be adjusted to yield exact solutions. We consider 
three individual cases in the following:

\subsection{Hyperbolic Potential: $cosh(2\phi)$}

In four dimensions, a form of potential
$V(\phi)\propto ({\cal H}(\phi))^n$, where ${\cal H}(\phi)$ is a hyperbolic function, either
``sinh'' or ``cosh'', and $n$ is a real number, has been discussed by a number of authors
in the context of massive dilaton black holes \cite{harvey, polwil} and cosmologies \cite{chimento}. We generalize our potential as $V(\phi)=\sum_n a_n{\cal H}^n$, a polynomial in ${\cal H}(\phi)$. 
Here we are in effect taking the advantages of the arbitrariness in $a_n$
in $V(\phi)$ and searching for exact solutions by adjusting them.
We first choose a  dilaton.  Since $V(\phi)$  is in terms of a hyperbolic function,
it is reasonable to set $\phi(r)$ to be a logarithmic function in order to
simplify the field equations.  We  assume that
\begin{equation}
\phi = {1\over 4}ln\left(1-{2B\over\sqrt{r^2+B^2}+B}\right) \label{dila}
\end{equation}
which is the choice made in four dimensions for asymptotically flat dilaton
black holes \cite{ghs, clh}. With this choice of $\phi(r)$, (\ref{eeomb}) yields
\begin{equation}
h = {r^2\over r^2+B^2}. \label{hfun} 
\end{equation}
Now using (\ref{dila}) and (\ref{hfun}) to simultaneously solve (\ref{eeoma}) and (\ref{eeomc}) for 
the coefficients $a_n$ and $f(r)$, we are unable to get exact solutions unless
$a_n=0$ ($n=0,2,3,4,...$), $a_1=2\Lambda$ and ${\cal H}=cosh(2\phi)$, that is,
\begin{equation}
V =  2\Lambda cosh(2\phi), \label{pota}
\end{equation}
and 
\begin{equation}
f  = \Lambda r{\sqrt{r^2+B^2}}. \label{1sola}
\end{equation}
Summing up, the Lagrangian and the corresponding solution are
\begin{equation}
{\cal L}={\cal R}-4(\nabla\phi)^2+2\Lambda cosh(2\phi), \label{1a}
\end{equation}
\begin{equation}
ds^2= -(\Lambda r{\sqrt{r^2+B^2}})dt^2+{rdr^2\over \Lambda (r^2+B^2)^{{3\over 2}}}
+r^2d\theta^2, \quad \phi = {1\over 4}ln\left(1-{2B\over\sqrt{r^2+B^2}+B}\right).\label{1b}
\end{equation}
The only curvature singularity is located at $r=0$ and it is naked and timelike.
In section 7, we will investigate
whether a quantum particle still behaves singularly or not when it approaches 
near the singularity.
Using (\ref{massb}), one can show that the mass parameter at $r\rightarrow\infty$
is $2M=-3\Lambda B^2$ for the choice of the background solution $M=0\Rightarrow W_{0}(r)=\sqrt{\Lambda} r$. One must have a negative mass solution
for the correct signature of the metric.
The solution (\ref{1b}) is for a non-vanishing $B$; note that
$B=0\Rightarrow\phi=0$ and $V=2\Lambda$. 
So $B=0$ yields a zero mass $BTZ$ solution.
As $r\rightarrow\infty$, $\phi\rightarrow 0$ and (\ref{1b}) asymptotically
approaches the massless $BTZ$ solution.
Changing the choice of $\phi(r)$ may lead to new solutions
but we are unable to get exact black hole solutions with the
above hyperbolic potential. 

\subsection{First Trigonometric Potential: $cot\phi$}

We now turn our attention to trigonometric functions in $V(\phi)$.
One example is the Sine-Gordon potential which is currently under investigation in two dimensional 
curved spacetime (see {\sl e.g.} \cite{sine}). We are unable
to get exact solutions for the Sine-Gordon potential either.
Rather than assuming $V(\phi)$
in terms of Sine-Gordon form, we generalize $V(\phi) = \sum_n b_n{\cal T}^n(\phi)$,
where ${\cal T}(\phi)$ is an unspecified trigonometric function. By following the approach similar  to that demonstrated above, we find that an exact solution exists when
${\cal T}=cot\phi$. The Lagrangian  and solution are
\begin{equation}
{\cal L} ={\cal R} - 4(\nabla\phi)^2 
+ 2\Lambda + 3\Lambda cot^2\phi + \left({4M\over 27B^2} - \Lambda\right)cot^6\phi,  \label{potb} 
\end{equation}
\begin{equation}
ds^2= -\left[-M\left(1 -  {B\over r}\right) + \Lambda r^2 \right ]dt^2+ 
{\left(1-{3B\over 2r}\right)^2\over -M\left(1 -  {B\over r}\right) + \Lambda r^2}dr^2 + r^2d\theta^2, \quad r  = {3B\over 2}sec^2(\phi). \label{esolb}
\end{equation}
We first comment on the status of constants, $B$ and $M$ which appear explicitly in the
Langrangian. ($\Lambda$ is the analogous cosmological constant.)
There are three possible situations. First, both  $B$ and $M$ are fixed coupling constants.  
Black holes for this action are like
discrete bound state solutions since their metrics depend only on coupling constants. Examples of such black holes are not uncommon
and can be found in two \cite{chanmann} and four \cite{bech, romans} dimensions
respectively. Second, both $B$ and $M$ are integration constants. 
In this case, it is only the combination ${M\over B^2}=b^2$ which appears in the Lagrangian (\ref{potb}) and which should be considered as a coupling constant. Now replacing $B={\sqrt{M}\over b}$ in the metric 
(\ref{esolb}), we can regard  $M$ as an  integration constant in the metric.  
When using (\ref{massb}) to compute the mass at spatial infinity when $\Lambda>0$ (no cosmological horizon) by setting $M=0$ in the background solution, one find that it is diverging.  The third case is to consider 
$M$ as an integration while $B$ as a coupling constant by introducing a three-form
field in the action. In section 6, we will illustrate how this can be done. 
Now using (\ref{massb}) one find that $M={\cal M}(\infty)$ is the finite mass parameter when the background $M=0$ is chosen. Although the metric is the same in the second and third situations, 
it is not asymptotically flat and therefore no unique background metric exists. Different choices of integration constants may lead to different backgrounds. As a result, one may get different values of mass for a given $r$. From now on, we will consider only the third situation. 

The dilaton in (\ref{esolb}) implies that $B>0$. Different relative signs and/or magnitudes between $B$ and $\Lambda$ may lead to solutions with or without
event/cosmological horizons. When $\Lambda=0$, the solution is asymptotically flat but $M$ shown in (\ref{esolb}) is no longer the total mass measured at spatial infinity. Instead,  one has to use (\ref{massb}) again to identify the total mass.  When $\Lambda=0$, we replace $M$ by $C$ in (\ref{esolb}). Using Minskowski metric as the background we find that the mass is given by $2(1-\sqrt{C})=M$.  When we further set $B=0$ and $\phi={\pi\over 2}$, then the metric reduces to the locally flat conical spacetime in three dimensions obtained in \cite{dhj}.

For the $\Lambda\neq 0$ solution, there are several properties:

\noindent
(i) As $r\rightarrow\infty$, $\phi\rightarrow{\pi\over 2}$ in (\ref{esolb}) and $V\rightarrow
2\Lambda$ in (\ref{potb}). The asymptotic form of the metric is like the $BTZ$ solution.
Solution (\ref{esolb}) indicates that when $B=0$, one must have
$\phi={\pi\over 2}$ and as a consequence $V=2\Lambda$. 
As a result, one gets back the 
$BTZ$ black hole solution. In that sense, the solution (\ref{esolb}) is a result from the modification of the  $BTZ$ black hole due to the dilaton given in (\ref{esolb}).

\noindent
(ii) There is a ``ring'' curvature
singularity at $r_s={3B\over 2}$. If $\Lambda>0$, then the range of $r$ is
$\infty > r > {3B\over 2}$. One can imagine that there is a special
collapsing dilaton fluid which collapses into a ring rather than a point
due to a strong enough outward radial pressure at $r_s$.
Note that a similar solution can be found in four dimension: The
string magnetic black hole admits a singular two-sphere \cite{ghs}.

\noindent
(iii) $-g_{tt}=0$ signals the location(s) of event/cosmological horizon(s).
Recall that both conditions $M<0$ and $B<0$ are disregarded.
A simple graphical analysis indicates that
there is only one turning point (a minima) for $-g_{tt}$ if 
$\Lambda>0$. The condition $\Lambda>0$ yields at most 
two real positive roots which are the outer and inner horizons. 
Since $-g_{tt}=0$ is a cubic equation, the roots are given by
\begin{equation}
r_k = -\sqrt{{4\over 3}{M\over\Lambda}}cos{\delta}_k,  \quad \&  \quad  
cos3{\delta}_k = {BM\sqrt{\Lambda}\over 2}\left({3\over M}\right)^{3\over 2}.
\label{root}
\end{equation}
The two real positive roots of (\ref{root}) are the
outer and inner horizons, $r_{\pm}$.
Concerning the $\pm$ sign of the horizon, from now on it is
understood that when there is only one event horizon the $+$ sign is chosen.
As long as $4M\geq 27B^{2}\Lambda$,
the outer and inner horizons do not vanish.
As long as $r_+$ (or $r_-$) is greater than
$r_s={3B\over 2}$, the ring curvature singularity is not naked.
The extremal condition is $4M=27B^{2}\Lambda$ and consequently the third term of corresponding potential function (in (\ref{potb})) vanishes.  
For the extremal case, using (\ref{root}), it is easy to see that $r_+=r_- =
\left(BM\over 2\Lambda\right)^{1\over 3}={3B\over 2}=r_s$. The outer horizon, inner horizon
and ring singularity coincide and the metric is given by
\begin{equation}
ds^2 = -{\Lambda\over r}(r+3B)\left(r-{3\over 2}B\right)^2dt^2 + {dr^2\over \Lambda r(r+3B)} + r^2d\theta^2, \quad B=2\sqrt{{M\over 27\Lambda}}. \label{extbh}
\end{equation}

\noindent (iv) We now consider the causal structure of the black hole
solution. Since $-g_{tt}$ in (\ref{esolb}) is cubic, we are unable to generally
integrate $g_{rr}$ to get the tortoise co-ordinate $r_*$. However,
we can still deduce what the black hole causal structures look like.
Note first that
(\ref{esolb}) asymptotically approaches the $BTZ$ metric, therefore the
null infinity is timelike. If $r_+>r_s>r_-$, then one may deduce that the
ring singularity is spacelike. Thus the causal structure of case
$r_+>r_s>r_-$ is like the static and uncharged $BTZ$ causal structure. 
We will not repeat the drawing here.
If $r_+>r_->r_s$, then the ring singularity would be  timelike
(every time we cross a horizon,
the space and time co-ordinates interchange roles).
As a result, the causal structure of case $r_+>r_->r_s$
looks like the spinning $BTZ$ black hole.
For the extremal case, $r_+=r_->r_s$, it looks like the extremal $BTZ$ case.

\subsection{Second Trigonometric Potential: $cot{\sqrt{2}\phi}$}

There is a second solution if we consider
both ${\cal T}=cot{\sqrt{2}}\phi$. The action and solutions are   
\begin{equation}
{\cal L} = {\cal R} - 4(\nabla\phi)^2 
+2\Lambda + 2\Lambda cot^2\sqrt{2}\phi
+ \left({M\over 2L}-2\Lambda\right) cot^4\sqrt{2}\phi +
  \left({M\over 2L}-2\Lambda\right) cot^6\sqrt{2}\phi, \label{potential}
\end{equation}
\begin{equation}
ds^2 = -\left[-M\left(1-{L\over r^2}\right)+\Lambda r^2\right]dt^2
+{\left(1-{2L\over r^2}\right)^2\over -M\left(1-{L\over r^2}\right)+\Lambda r^2}dr^2+r^2d\theta^2, 
\quad r^2 =  {2L}sec^2(\sqrt{2}\phi). \label{sola}
\end{equation}
$M={\cal M}(\infty)$ is the mass parameter.  
As before, we only consider $M>0$
and the dilaton shown in (\ref{sola}) implies $L>0$.
The properties of this solution are very similar to the previous
one. Again there is a ring curvature singularity located at $r_s^2={2L}$.
The range of $r^2$ is therefore $\infty>r^2>{2L}$ if $\Lambda>0$.
When $L=0$, (\ref{sola}) reduces to the BTZ case. 
One must have $\Lambda>0$ for the existence of an
event horizon(s). The location of the outer and inner horizons are given by
\begin{equation}
r_{\pm}^2 = {M\pm\sqrt{M^2-4\Lambda LM}\over 2\Lambda}. \label{roots}
\end{equation}
It can be checked that
$r_{+}^2$ ($r_-^2$) is always greater (less) than $r_s^2$. Thus the inner horizon is
physically irrelevant.  For the extremal case,
$r_+^2 = r_-^2 = r_s^2={2L}={M\over 2\Lambda}$;  all horizons coincide together and the third and fourth terms of the corresponding potential function in (\ref{potential}) vanish. 
The causal structures of (\ref{sola}) can be deduced
in a manner similar to (\ref{esolb}). We will not repeat the discussion here.

\section{Exact solutions in $ST$ gravity}

We now look for solutions in $ST$ gravity described
by field equations (\ref{eqsta}) and (\ref{eqstb}). Similar to the dilaton gravity cases studied in previous section, we consider a potential function which was previously
investigated in four dimensional $ST$ gravity.
We note that a potential of the form $\sum_n c_n\phi^n$ (a polynomial of
degree $n=4$) was discussed
in \cite{madsen} for a special class of $ST$ gravity in four dimensions.
Here we again generalize $V$ to a polynomial: 
$\sum_n c_n\phi^n$. We will adjust  the coefficients $c_n$ 
to get exact solutions. Given this form of $V(\phi)$, we find that there exist three exact solutions correspond to three different choices of scalar.

\subsection{Case i: $\phi = r/(r-3B/2)$}

The first example corresponds to the choice
\begin{equation}
\phi = {r\over r-{3B\over 2}}. \label{edilc}
\end{equation}
Now (\ref{eeqstb}) yields the coupling function
\begin{equation}
\omega = {2\over 1-\phi}, \label{ecfunc}
\end{equation}
By following the similar procedures outlined above, the action and solution are
\begin{equation}
{\cal L} = \phi{\cal R}-{2\over 1-\phi}(\nabla\phi)^2
+ 2(3 - 3\phi + \phi^2)\Lambda\phi + {8M\over 27B^2}(1-\phi)^3,  \label{epotc}
\end{equation}
\begin{equation}
ds^2 = -\left[-M\left(1 - {B\over r}\right) + \Lambda r^2\right]dt^2+{dr^2\over -M\left(1 - {B\over r}\right) + \Lambda r^2} + r^2d\theta^2, \quad \phi = {r\over r-{3B\over 2}} . \label{esolc}
\end{equation}
It can be checked that (\ref{eqstb}) is satisfied as well.
$M$ is the mass parameter calculated using (\ref{massc}) in the background
$W_{0}=W(M=0)$ and the limit $r\rightarrow\infty$.
We require that $M>0$. 

Unlike the dilaton case, $B$ is no longer required to be positive.
The metric looks like the Schwarzschild-(anti)-de Sitter metric except
for the position of the mass term. If $\Lambda=0$, the metric is exactly the
same form as the four dimensional Schwarzschild case. 
There are several properties of metric (\ref{esolc}):

\noindent
(i) As $r\rightarrow\infty$, $\phi\rightarrow 1$, $V(\phi)\rightarrow 2\Lambda$
and $\omega(\phi)\rightarrow\infty$, the metric asymptotically approaches
the $BTZ$ metric. Obviously, if $B=0$, then $\phi=1$ and the metric is
exactly the $BTZ$ black hole. The $BTZ$ black hole is therefore modified by
the scalar in (\ref{edilc}).

\noindent
(ii) Although ${\cal R}=-6\Lambda$ (constant curvature scalar),
the Krestmann scalar diverges as $r\rightarrow 0$. The only curvature
singularity is located at $r=0$. Note that unlike the dilaton case, $r={3B\over 2}>0$
does not correspond to a curvature singularity but rather to a diverging $\phi(r)$
and $V(\phi)$ in (\ref{edilc}) and (\ref{epotc}). This situation is similar to the situation in the
four dimensional Bekenstein black hole \cite{bek}, where the conformal coupling
scalar in the action diverges at the event horizon, and to the situation in
several classes of two dimensional black holes \cite{chanmann}, where the kinetic
term of a scalar field in the action
diverges at the outermost event horizon. In \cite{chanmann, bek}
the geometry is still well-behaved at the event horizon.
In the present case, the geometry is well-behaved at
$r={3B\over 2}$, although a non-freely falling
test particle coupling to $\phi(r)$ and traveling towards the event horizon
from infinity might encounter a diverging barrier at this value of $r$.
We will not consider this as a disaster since the existence of such a barrier solely
depends on how the non-freely falling test particle couples to $\phi$, and
there is no unique way of determining what this coupling should be (see discussions
in \cite{chanmann}). The only restriction we impose is $r_+\neq {3B\over 2}$ in order to avoid a diverging entropy $S$, since $S\propto\phi(r_+)$ (see next section).

\noindent
(iii) An interesting situation is given by the conditions: $B>0$, $\Lambda>0$ and $4M\geq 27B^{2}\Lambda$, in which 
the metric (\ref{esolc})  generally admits an outer and an inner horizon.
This example shows that the existence of an inner horizon is not
necessarily due to the presence of an electric charge/angular momentum.
If the equality sign holds for the latter, the outer and inner horizons
will coincide, $r_+=r_-={3B\over 2}$. One may expect a diverging entropy for this extremal black hole. However, an extremal black hole need a separate treatment in thermodynamics. 

\noindent (iv) A special  case of interest is given by the condition: $B=-2\sqrt{{M\over 27\Lambda}}$. This condition yields  $V=2\Lambda$ in (\ref{epotc}). 
If we carry out a scalar redefinition $\phi = 1-{1\over 8}\Psi^2$, the 
corresponding theory and solution are 
\begin{equation}
S=\int  d^3x {\sqrt{-g}} [\left(1-{1\over 8}\Psi^2\right){\cal R}-(\nabla\Psi)^2
+ 2\Lambda], \label{ctheory}
\end{equation}
\begin{equation}
ds^2 = -F(r)dt^2 + {dr^2\over F(r)}+r^2d\theta^2, \quad F={\Lambda}\left[r^2-{M\over \Lambda}-{2\over r}\left({M\over 3\Lambda}\right)^{{3\over 2}}\right],  \quad  
\Psi=\sqrt{8}\sqrt{{\sqrt{{M\over 3\Lambda}}\over r+\sqrt{{M\over 3\Lambda}}}}.
\label{cbh}
\end{equation}
The event horizon is located at 
\begin{equation}
r_+=2\sqrt{{M\over 3\Lambda}}. \label{r_+}           
\end{equation}
The above action and solution is exactly the conformal black hole  obtained in \cite{ZAN} by Martinez and Zanelli\footnote{Note that  the mass parameter, ${\tilde M}$,  in \cite{ZAN} is one-eighth of $M$ shown in (\ref{cbh}). That is,  $8{\tilde M}=M$. It is because the authors defined the entropy in general relativity as one-fourth of the area of the event horizon, while we use the definition \cite{btz, broyor}  that in three-dimensional general relativity, the  entropy is twice the area.}.  Starting from action (\ref{ctheory}), the authors solved the field equations directly to get (\ref{cbh}). 

\noindent (v) 
Note  that the coupling function (\ref{ecfunc}) is a function of $r$. It
has different signs for different ranges of $r$. For example,
$\omega(\phi)>0$ for $B>0$ and $r<{3B\over 2}$, and $\omega(\phi)<0$
for $r>{3B\over 2}$. In some sense the kinetic term flips sign.
This situation is very similar to the cases in \cite{chanmann}.
The physics of the transition $\omega>0\leftrightarrow\omega<0$ and the stability
of the above solution are open questions to be investigated.

\subsection{Case ii: $\phi=2/(2+kr)$}

By assuming a different form of scalar, we get
\begin{equation}
{\cal L} = \phi{\cal R}+{2\over \phi}(\nabla\phi)^2
+2\left(\Lambda - {Ak^2\over 4}\right)\phi^3, \label{epotd}
\end{equation}
\begin{equation}
ds^2=-[A(1 + kr) + \Lambda r^2]dt^2+{dr^2\over A(1 + kr) + \Lambda r^2} + r^2d\theta^2, \quad 
\phi = {2\over 2 + kr}. \label{edild}
\end{equation}
This metric is exactly the same form as the one obtained in a special class
of two dimensional $ST$ gravity \cite{mann}. It is easy to show that the mass $M\rightarrow 0$ as
$r\rightarrow\infty$ in (\ref{edild}) if $k\neq 0$. Thus the analogous $ADM$
mass vanishes. This ``massless'' black hole is in fact a Brans-Dicke type black hole
with a non-vanishing
potential. If $k=0$, $A$ is identified as the negative mass parameter, that is to say,
$A=-M$, and the usual $BTZ$ metric is recovered.
Note that a similar form of metric (\ref{edild}) 
was previously obtained in \cite{sa} in three dimensional Brans-Dicke
theory with a different choice of $\phi(r)$, and with $V(\phi)=0$
and $\omega=-{2\over \phi}$. 

The potential function in (\ref{epotd}) and scalar in (\ref{edild}) both diverge at $r=-{2\over k}$.
The existence of the event horizon(s) can easily be investigated
by graphical analysis. It is interesting to note that for the extremal
case, $Ak^2=4\Lambda$, the potential function vanishes.

\subsection{Case iii: $\phi=r^2/(r^2-2L)$}

Finally,  a third choice of an asymptotically constant scalar yields
\begin{equation}
{\cal L} = \phi{\cal R}-{4\phi-1\over 2\phi(1-\phi)}(\nabla\phi)^2
+ {M\over 2L} + 6\left(2\Lambda-{M\over 2 L}\right)\phi
+18 \left(-\Lambda+{M\over 4L}\right)\phi^2
+ 2\left(4\Lambda - {M\over L}\right)\phi^3,  \label{epote}
\end{equation}
\begin{equation}
ds^2 =  -\left[-M\left(1-{L\over r^2}\right)+\Lambda r^2\right]dt^2+{dr^2\over -M\left(1-{L\over r^2}\right)+\Lambda r^2} + r^2d\theta^2, \quad  \phi = {r^2\over r^2-2L}. \label{esole}
\end{equation}
$M$ is the mass parameter.
The metric has a curvature singularity at $r=0$.
The scalar and its potential both diverge at $r^2={2L}$ with $L>0$.
The extremal limit is $4\Lambda L=M$. In this limit, the third term in the Lagrangian becomes $2\Lambda$, while fourth, fifth and sixth terms all vanish.  
The properties of (\ref{esole}) are very similar to the previous cases and we do not repeat the discussions here.
Note that one can generalize the three scalars in  (\ref{edilc}), (\ref{edild}), and (\ref{esole}) to
$\phi={r^i\over C_1r^j+C_2}$, where $i$ and $j$ are integers. However, we are unable 
to get exact solutions for arbitrary $i$ and $j$.   

Before ending this section, we note two things. 
First,  a similar {\sl ad-hoc}
procedure was adopted by Bechmann and Lechtenfeld \cite{bech}
to study the modifications of the four dimensional Schwarzschild black hole
in general relativity.
They reversed the role of the dilaton potential $V(\phi)$, keeping the
choice of $V(\phi)$ initially unspecified and then solving it
by using a desired $\phi(r)$ and an asymptotically form of the metric.
As a result, the Schwarzschild mass parameter
appears in the potential explicitly. In this and the previous section, we have done
a similar thing:
We studied how a dilaton or scalar modifies the three
dimensional $BTZ$ black hole in general relativity. We use the freedom in the coefficients of
the potential and coupling functions to construct exact solutions.
Those potential forms are not completely arbitrary since they were previously
studied in other dimensions. Second,  note that the above dilaton and $ST$
solutions are obviously not unique to the field equations followed from (\ref{act}):
changing the choices
of those dilatons, scalars and coefficients and/or the forms of those potential
and coupling functions can lead to other solutions. They may or may
not be the modification of $BTZ$ solutions. Here we just illustrate
some possible solutions which are modifications of the
$BTZ$ case. In the next section, we will briefly discuss how the mass parameter in the actions be absorbed by a three-form and some of the thermodynamic  properties of the solutions.

\section{Temperature}

In all of the above black hole solutions, the analogous $ADM$
mass parameter $M={\cal M}(\infty)$ identified
at spatial infinity through (\ref{massb}) and (\ref{massc}) appears
as a universal constant in the 
actions. In this sense, black holes exist only with specific
masses determined by some of the parameters in the actions. Black holes for these actions are thus rather like discrete bound state solutions.
This is not  of  much  interest since one can only consider one solution
of the field equations at a time. It is much more satisfactory to regard $M$ as an
integration constant rather than a universal constant in the action.
In order to treat $M$ as a dynamical variable, one must be able to
vary it in the action. This is impossible for the above black hole
solutions since $M$ appears explicitly in the potential functions.
A way is needed to restore $M$ to its status as an integration constant
in such a way as to yield the same effective field equations and solutions.

Similar to the method used in four dimensions \cite{jolien, teit}
where the cosmological
constant is treated as a dynamical variable by coupling  gravity to a four-form
field, we will couple gravity to a gauge invariant
three-form field $H_{\mu\nu\rho}$,
where $H_{\mu\nu\rho}=\nabla_{[\mu}A_{\nu\rho]}$
and $A_{\mu\nu}$ is a completely antisymmetric two-index gauge field.
Note that three-form fields arise naturally in low energy
string gravity in any dimension, except in two dimensions where any three-form vanishes.
However, one can also regard the $H_{\mu\nu\rho}$ as a possible matter field
to any non-string theory of gravity.
Here our emphasis is on how a three-form matter field
can introduce a dynamical variable to three dimensional dilaton
and $ST$ gravitational theories.
We only discuss $H_{\mu\nu\rho}$-coupling in the dilaton
gravity cases, since the generalization of it
to $ST$ gravity is straightforward. 

The generalized action for such a three-form coupled to dilaton gravity can be
written as \cite{jolien}
\begin{equation}
S = \int  d^3x\sqrt{-g}({\cal R}-4(\nabla\phi)^2+V(\phi)+l(\phi)H_{\mu\nu\rho}H^{\mu\nu\rho}), \label{acta}
\end{equation}
where $l(\phi)$ is another coupling function.  The $H_{\mu\nu\rho}$
equation of motion is given by
\begin{equation}
\nabla^{\mu}(l(\phi)H_{\mu\nu\rho}) = 0, \label{hmotion}
\end{equation}
and its solution is
\begin{equation}
H_{\mu\nu\rho} = {q\over l(\phi)}\epsilon_{\mu\nu\rho}, \label{hsol}
\end{equation}
where ${\bf\epsilon}$ is the volume three-form, and $q$ is an integration
constant (dynamical variable) which is proportional to
the conserved charge (analogous to the electric charge for the two-form Maxwell field)
associated with the gauge field $A_{\mu\nu}$.
Mathematically $V(\phi)$ is a 0-form, $\partial_{\mu}\phi$ is a 1-form and $H_{\mu\nu\rho}$ is a 3-form. Thus action
(\ref{acta}) accommodates $n$-form ($n=0, 1, 3$) fields in three dimensional dilaton gravity.

It is not hard to check that the last term in action (\ref{acta})
produces an effective potential term: $V_{eff}={6q^2\over l(\phi)}$
in the right hand side of ${\cal R}_{\mu\nu}$ in (\ref{eoma})
and ${dV_{eff}\over d\phi}$
in the left hand side of (\ref{eomb}). For the Lagrangian (\ref{potb}) of the
black hole solution (\ref{esolb}), we write the effective potential term as,
\begin{equation}
V_{eff} = {4M\over 27B^2} cot^6\phi. \label{hpot}
\end{equation}
by setting $q^2={M}$ and
$l^{-1}(\phi)={2\over 81}B^2cot^6\phi$. Thus for a fixed $B$, $M$ can now be treated
as an integration constant (dynamical variable).
For the second dilaton black hole solution (\ref{sola}), we can write
\begin{equation}
V_{eff} = {M\over 2L}\left(cot^4\sqrt{2}\phi+cot^6\sqrt{2}\phi\right) \label{hpota}
\end{equation}
by setting $q^2={M}$ and
$l^{-1}(\phi)={L\over 12}(cot^4\sqrt{2}\phi+cot^6\sqrt{2}\phi)$.
To sum up, by using the generalized $n$-form action (\ref{acta}),
the mass parameter of the dilaton black hole solutions to the effective field equations
of (\ref{acta}) is now an integration constant instead of universal constant.
One can also add a similar three-form matter field to the $ST$ gravity to represent the
part(s) of the potential functions that have mass dependence.

We first compute the entropy of the dilaton black hole
solutions. It is now well recognized that non-extremal black hole entropy $S$
depends only on the geometry of the horizon. General arguments like those
in four dimensions \cite{jolien,visser} show that
\begin{equation}
S = 4\pi r_+ \label{enta}
\end{equation}
in three dimensional stationary black holes in Einstein gravity
coupled to ``dirty matter'' ({\sl e.g.} dilaton and generalized n-form
gauged fields), where $r_+$ is the outermost event horizon given by (\ref{root})
and (\ref{roots}).  The temperature 
is generally given by 
\begin{equation}
T={\kappa_H\over 2\pi}{1\over N},  \label{qtemp}
\end{equation}
where $N(r)$ is the lapse function for a stationary black hole spacetime, and $\kappa_H$ is the surface gravity \cite{jolien, joliena}. 
For an asymptotically non-flat spacetime, $N(r)$ does not approach 1 
as $r\rightarrow\infty$. In fact, ${\kappa_H\over 2\pi}$ is just
the temperature at the spatial location where $N(r)=1$,
not the Hawking temperature. For an asymptotically flat black hole,
one recovers the usual Hawking temperature
$T_H={\kappa_H\over 2\pi}$ since $N(\infty)\rightarrow 1$.
We only show the surface gravity in the following. 
For the black hole metric (\ref{esolb}), the temperature is
\begin{equation}
T = {-MB+2\Lambda r_+^3\over 4\pi r_+\left(r_+-{3B\over 2}\right)}. \label{tempa}
\end{equation}
It is easy to check that when $B=0$, the above expression reduces to
the $BTZ$ case studied by Brown, Creighton and Mann in \cite{broyor}.
Recall that for the extremal black hole, one has
$r^3_+={MB\over 2\Lambda}=\left ({3B\over 2}\right )^3$
and therefore $T_{ext}={1\over 4\pi r_+}$ which is non-vanishing. 
For the second dilaton
black hole (\ref{sola}), its temperature  is 
\begin{equation}
T = {(-ML+\Lambda r_+^4)\over 2\pi r_+\left(r_+^2-{2L}\right)}. \label{tempb}
\end{equation}
When $L=0$ the temperature reduces to the $BTZ$ case.  $T={1\over 2\pi r_+}$ for the extremal case.

We now turn to the $ST$ black hole solutions. Due to the presence of $\phi$
in the Ricci scalar in action (\ref{act}), the entropy is modified to be \cite{jolien, joliena}
\begin{equation}
S = 4\pi r_+ \phi(r_+). \label{entmod}
\end{equation}
Thus for black holes (\ref{esolc})  and (\ref{esole}), the entropies are
\begin{equation}
S = {4\pi r_+^2\over r_+-{3B\over 2}}, \label{entb}
\end{equation}
and
\begin{equation}
S = {4\pi r_+^3\over r_+^2-2L}. \label{entc}
\end{equation}
respectively. Note that in both solutions the entropy is ill-defined in the extremal cases where $r_+={3B\over 2}$ and  $r_+^2=2L$ respectively.  The temperature is still given by the surface gravity divided by the lapse \cite{joliena}.
For (\ref{esolc}), it is given by
\begin{equation}
T = {1\over 4\pi}\left(-{MB\over r_+^2}+2\Lambda r_+\right), \label{tempc}
\end{equation}
and for (\ref{esole}), it becomes
\begin{equation}
T = {1\over 2\pi}\left(-{ML\over r_+^3}+\Lambda r_+\right). \label{tempd}
\end{equation}
In both cases, $T$  vanishes in the extremal limits.
In addition, they reduce to the $BTZ$ temperature when $B=0$ and $L=0$ respectively.
In particular, for the $MZ$ solution (\ref{cbh}), recall that one has  $B=-2\sqrt{{M\over 27\Lambda}}$. Using 
(\ref{entb}) and (\ref{tempc}), we get $S={8\pi\over 3}r_+$ and $T={9\Lambda\over 16\pi}r_+$.
Equation (\ref{r_+}) implies that $M={3\Lambda\over 4}r_+^2$. Now it is easy to check that $dM=TdS$ as shown in \cite{ZAN}.  

Note that using the formalisms discussed in \cite{jolien, joliena}, one may compute various thermodynamics quantities for all the black hole solutions in this paper and study in detail their thermodynamics\footnote{For example, one may study the first law, $dE=TdS+\cdot\cdot\cdot(workterms)$, where $E$ is the quasilocal energy. Note that one should let the asymptotic value of dilaton/scalar to be arbitrary, $\phi(\infty)\rightarrow\phi_o$, since $\phi_o$ is considered as one of the thermodynamics variables \cite{CCM}.}.

\section{Quantum Probe of Singularity}

In general relativity, a spacetime is said to be singular
if it is (timelike/null/spacelike) geodesically incomplete.
In particular, if a spacetime admits a timelike singularity,
then the evolution of a test particle is not defined after a finite proper time.
Using the dilaton solutions of this paper as examples, 
we will investigate whether they are examples of static spacetimes which admit timelike
singularities in which the evolution of
a quantum test particle is completely well-defined near the singular
regions of the classical spacetimes.

Note that if the spacetime is free of any timelike singularity
and globally hyperbolic, then the evolution of
a quantum test particle must be essentially well-defined.
We will adopt the point of view \cite{hm} that a quantum particle
is non-singular when the evolution
of any state is uniquely defined for all time. If this is not the case,
then there is some loss of predictability and we will say that the particle
is singular. The existence of a non-singular quantum particle can be
translated into the existence of the self-adjointness of the spatial part
of the wave operator \cite{hm, walda}. A relativistic
spinless particle with mass $m\geq 0$
is described quantum mechanically by a positive frequency solution to the
wave equation of mass $m$. That is, the Klein-Gordon equation in curved
spacetime
\begin{equation}
\nabla^2\psi-m^2\psi = 0, \label{kg}
\end{equation}
where $\psi$ is the wave function of the particle. For a static spacetime with
a timelike Killing vector field $\xi^{\mu}$ with Killing parameter
$t$ (the time co-ordinate $x^0$), the wave equation (\ref{kg}) can be rewritten
in the form
\begin{equation}
{\partial^2\psi\over\partial t^2} = \Xi D^i(\Xi D_i\psi)-\Xi^2m^2\psi, \label{kga}
\end{equation}
where $\Xi^2=-\xi^{\mu}\xi_{\mu}=-g_{tt}$, and $D_i$ ($i=1,2$) is the spatial
covariant derivative on $\Sigma$, a static slice. Let ${\cal A}$ denote (minus)
the operator on the right hand side
\begin{equation}
{\cal A} = -\Xi D^i(\Xi D_i) + \Xi^2m^2. \label{oper}
\end{equation}
The question is then whether ${\cal A}$ is essentially self-adjoint for a spacetime
which admits a timelike singularity. It is our adopted criterion for defining
the quantum regularity/singularity. The meaning of self-adjointness
(see \cite{hm}) can be summarized as follow: Consider the equation ${\cal A}\psi\pm i\psi=0$
($i=\sqrt{-1}$). Using the separation of variables $\psi=Y(r)\Phi(\theta)$
and metric (\ref{escs}), one gets the following radial equation for $Y$
\begin{equation}
Y''+\left({f'\over f} - {h'\over 2h} + {1\over r}\right)Y'
- \left({c^2h\over fr^2} + {m^2h\over f} \pm i{h\over f^2}\right)Y = 0. \label{ey}
\end{equation}
$c^2$ is an eigenvalue of (minus) the Laplacian on the $1$-sphere.
The operator ${\cal A}$ will be essentially self-adjoint in the neighbourhood
of a radial co-ordinate if one of the two solutions (for each constant $c^2$ and
each sign of the complex term) to (\ref{ey}) fails to be square integrable with respect
to the measure ${r\sqrt{h}\over \Xi^2}={r\sqrt{h}\over f}$,
that is if 
\begin{equation}
I = \int{YY^{*}{r\sqrt{h}\over f}dr} \label{integration}
\end{equation}
diverges in the neighbourhood of the radial point of question \cite{hm}.
Here we will take the radial point near the timelike singularity.
Define a new radial co-ordinate $dr_*=g_{rr}dr$ so that the radial null
geodesics follows curves of constant $t\pm r_*$. If the singularity is
at a finite $r_*$, then it is timelike.

Since there are many solutions with naked timelike singularities
in three spacetime dimensions, we plan not to investigate them all.
Our intention in this section is to just illustrate the idea by
using several examples in dilaton cases.
We first investigate the naked solution (\ref{1b}). The curvature singularity is
located at $r=0$ and the finiteness of $r_*$ at $r=0$ indicates that
it is timelike. As $r\rightarrow 0$, the $m^2$ and complex terms in (\ref{ey})
may be ignored. In addition, ${f'\over f}\rightarrow{1\over r}$
and ${h'\over 2h}\rightarrow {1\over r}$. Thus in the neighbourhood of
$r=0$, (\ref{ey}) becomes $rY''+Y'-{c^2\over\Lambda B^3}Y=0$
which has the solution
$Y=C_1I_0(\alpha\sqrt{r})+C_2K_0(\alpha\sqrt{r})$, where $C_1$ and $C_2$ are arbitrary constants, $\alpha={2c\over\sqrt{\Lambda B^3}}$,
and $I_0$ and $K_0$ are the modified Bessel functions of first and second kind of zero order respectively. 
$K_0\rightarrow ln(r)$ and $I_0\rightarrow 1$ as $r\rightarrow 0$. Therefore  it is easy to see that (\ref{integration}) is finite near $r=0$. Thus ${{\cal A}}$ is not self-adjoint near $r=0$.
Previous examples of spacetimes with timelike singularities
which lead to singular quantum particles can be found in four dimensional
negative mass Schwarzschild solution and Reissner-Nordstrom solution \cite{hm}.

Next we consider solution (\ref{esolb}).
We set $M=0$ and choose $\Lambda>0$. Now the ring curvature
singularity at $r_s={3B\over 2}$ is naked and timelike. Note that
$f(r_s)\neq 0$ but $h(r_s)=0$. As $r\rightarrow r_s$, the $c^2$, $m^2$ and
complex terms in (\ref{ey}) can be ignored. Only the ${h'\over 2h}$ term is dominating.
As a result, one has $Y'\propto \sqrt{h}$. If we define $\rho=2r-3B$, then $\rho=0\Leftrightarrow r=r_s$ and near
$\rho=0$, the only form of solution is $Y= C_1\rho+C_2ln(\rho+3B)$. It can be checked that
$I$ in (\ref{integration}) is finite and ${\cal A}$ is not self-adjoint.
The final example is the solution (\ref{sola}). Similar to the
previous case, we set $M=0$. We further set $\rho^2=r^2-2L$
and $\rho=0$ is the location of the ring curvature singularity.
The repetition of the previous analysis indicates that ${\cal A}$ is
not self adjoint near $\rho=0$. Thus all three examples above lead to
singular quantum test particles.
It is worthwhile  mentioning that the static and chargeless $BTZ$ case with $M<0$.
It has no curvature timelike singularity at $r=0$. It is easy to check that
near $r=0$, one solution has the form $Y\propto r^{-{c\over \sqrt{-M}}}$
which can lead to a diverging $I$ near $r=0$. Thus the operator ${\cal A}$ is 
self-adjoint as expected. 

One can proceed similarly in the $ST$ gravity cases,
but with a different form of ${\cal A}$, which is
modified to include the coupling of the scalar $\phi$ to
quantum test particles (see \cite{hm}). We intend not to discuss them further.
It is worthwhile to mention that there are
no generic theorems to show that a quantum test particle
must be non-singular (in the sense of self adjointness of ${{\cal A}}$)
in a static spacetime with a timelike singularity. Thus the examples showed above
do not lead to any problem (it is indeed interesting to understand
under what conditions a quantum particle will behave non-singularly
in a singular spacetime). However, when approaching the strong field regime, the 
action (\ref{act}) might need to be modified (by properly introducing several extra dilaton/scalar fields) due to quantum gravitational effects.
Then one may use the modified solutions to investigate the singularity/regularity
of a quantum test particle. In the above solutions, we just assume that they are
all still valid in the neighbourhood the the singularities.

\section{Conclusion}

We have found five black hole solutions to
three dimensional dilaton and $ST$ gravity. 
When there are no cosmological horizons ($\Lambda>0$),
four of them asymptotically approach the $BTZ$ metric at the spatial infinity.
The forms of potential functions, which are polynomials,
are motivated from
their two/four dimensional counterparts. We leave the coefficients
of the polynomials, and the forms of coupling functions unspecified.
Here we took the advantage of this arbitrariness
to search for exact solutions. In the cases of dilaton gravity, the standard vacuum theory is general relativity and we search for certain dilaton matter actions in which their solutions contain the $BTZ$ spacetime as a special case. In the cases of $ST$ gravity, we construct vacuum theories which yield similar solutions. Once  particular coupling and potential functions have been determined,
one can regard the action containing these functions as a separate matter/theory in its own right and further explore the solution space of its field equations. 
Obviously, the action is a primary quantity. It is more interesting to know in advance what actions one can use and then extracting the corresponding solutions, than to construct the actions from solutions.  

In the dilaton gravity,  three different forms of potential  yield
three solutions respectively. Two of them are black holes with a ring curvature
singularity at a finite proper circumference which is hidden by the outer horizon. 
In the $ST$ gravity, three different choices of coefficient for a given form of potential yield three different black hole solutions. One solution contains the   Martinez 
and Zanelli conformal black hole as a special case.  In contrast to the dilaton cases,
none of them have any ring singularity. The only curvature singularity
is located at $r=0$ in each case. 

In all of the solutions, we use
the quasilocal formalism to identify the analogous $ADM$ mass parameter
$M$. We find that black holes exist only for
their mass $M$ determined by parameters in the potentials, and so
$M$ appeared as a universal constant in the action. However, by including a
coupling of dilaton or $ST$ gravity to a three-form field yields an
effective potential term in the field equations. By setting $M$ 
to be proportional to the charge of the three-form field, its status
as a constant of integration is restored.
We  calculate the temperature and entropy of the black hole solutions. 
They also provide examples to illustrate how one
can investigate whether a quantum particle behaves non-singularly or not in the
sense of self-adjointness of the operator ${\cal A}$. 

It is interesting to see whether one can also add angular momentum to
the present dilaton and $ST$ solutions, although the work is non-trivial.  Our present solutions depend on the choice of an asymptotically
constant $\phi(r)$, the coefficients of the potentials and the coupling functions.
By considering other choices of $\phi(r)$ and potential/coupling functions, one may extract more solutions which may or may not be containing the $BTZ$ as special cases.  
Also $\phi(r)$ is not necessarily required to be asymptotically constant, and
examples of such black hole solutions can be found in three \cite{chaman, spinchan} and four dimensions \cite{jhorne}. We hope that our
paper suggests a way for the derivation of more generic and
interesting solutions.  Recently,  Cai and Zhao \cite{cai} used the inner (Cauchy) horizon of a class of $(2+1)$-dimensional static charged dilaton black holes obtained in \cite{chaman} to test the conjecture of Cauchy horizon stability suggested by Helliwell and Konkowski. The conjecture is found to be valid. Similar investigations were carried out in \cite{su} for the spinning $BTZ$ black hole. 
Similarly, one can check the conjecture using the inner (Cauchy) horizon of the $ST$ black holes obtained in this paper. We intend to relate further details elsewhere.

\bigskip
\centerline{\bf Acknowledgements}

This work was supported in part by the Natural Sciences and Engineering
Research Council of Canada. The author would like to thank Robert Mann and Robert Myers for useful comments, as well as Jolien Creighton for discussions on the quasilocal formalism.

\end{document}